\documentclass{article}
\usepackage{spconf,amsmath,graphicx}
\usepackage{booktabs}
\usepackage{graphicx}
\usepackage{color}
\usepackage{booktabs}       %
\usepackage{multirow,array}
\usepackage{bm}
\usepackage{hyperref}

\usepackage{bm}
\usepackage{amssymb,amsmath,graphicx,bbm,color,booktabs}
\usepackage{amsopn}

\newcommand{\vech}{\bm{h}}

\newcommand{\vn}{\bm{n}}

\newcommand{\vx}{\bm{x}}       \newcommand{\vxh}{\hat{\bm{x}}}        
\newcommand{\vy}{\bm{y}}               
\newcommand{\vz}{\bm{z}}

  \newcommand{\Sc}{\mathcal{S}}

    \newcommand{\Xc}{\mathcal{X}}

\newcommand{\R}{\mathbb{R}}

\newcommand{\X}{\mathcal{X}}

\renewcommand{\eqref}[1]{Eq.~(\ref{#1})}

\title{High Fidelity Speech Regeneration \\with application to speech enhancement}

\name{Adam Polyak$^{1,2}$\sthanks{ \hspace{0.1cm} The contribution of Adam Polyak is part of a Ph.D. thesis
research conducted at Tel Aviv University.}, Lior Wolf$^{1,2}$, Yossi Adi$^{1}$, Ori Kabeli$^{1}$, Yaniv Taigman$^{1}$}
\address{
  $^1$Facebook AI Research \\
  $^2$The School of Computer Science, Tel Aviv University}

\usepackage{xcolor}

\begin{document}
\maketitle
\begin{abstract}
Speech enhancement has seen great improvement in recent years mainly through contributions in denoising, speaker separation, and dereverberation methods that mostly deal with environmental effects on vocal audio. To enhance speech beyond the limitations of the original signal, we take a regeneration approach, in which we recreate the speech from its essence, including the semi-recognized speech, prosody features, and identity. We propose a wav-to-wav generative model for speech that can generate 24khz speech in a real-time manner and which utilizes a compact speech representation, composed of ASR and identity features, to achieve a higher level of intelligibility. Inspired by voice conversion methods, we train to augment the speech characteristics while preserving the identity of the source using an auxiliary identity network. Perceptual acoustic metrics and subjective tests show that the method obtains valuable improvements over recent baselines.
\end{abstract}
\begin{keywords}
speech enhancement, audio generation
\end{keywords}
\section{Introduction}
\label{sec:intro}
Speech is the primary means of human communication. The importance of enhancing speech audio for better communication and collaboration has increased substantially amid the COVID-19 pandemic due to the need for physical distancing. 

In the domain of speech enhancement, denoising and dereverberation, methods have received much of the attention. The vast majority of these methods deal with environmental effects and train %
masking filters in order to remove unwanted sources, while assuming that the existing vocals are intelligible enough. However, in common environments where the recorded speech comes from a low fidelity microphone or poorly treated acoustic spaces, such methods struggle to reconstruct a clear sounding natural voice, which is similar to a voice that is recorded in a professional studio. 

Speech recognition and generation have seen remarkable progress in recent years mainly due to advances in the robustness of neural-based Automatic Speech Recognition (ASR) and neural vocoders. We utilize these advances and introduce a speech regeneration pipeline, in which speech is encoded at the semantic level through ASR, and a speech synthesizer is used to produce an output that is not only cleaner than the input, but also has better perceptual metrics.

Our main contributions are: (i) We present a novel generative model that utilizes ASR and identity information in order to recreate speech in high-fidelity through comprehension; (ii) We present quantitative and subjective evaluation in the application of speech enhancement, and; (iii) We provide engineering details on how to implement our method efficiently to utilize it in real-time communication. Samples can be found at {\color{purple} \url{https://speech-regeneration.github.io}}.

\section{Related work}
\label{sec:Related}
Speech enhancement and speech dereverberation were widely explored over the years~\cite{benesty2009noise,attias2001speech}. Due to the success of deep networks, there has been a growing interest in deep learning-based methods for both speech enhancement and dereverberation working on either frequency domain or directly on the raw waveform~\cite{stoller2018wave,su2020hifi,defossez2020real}. Deep generative models such as Generative Adversarial Network (GAN) or WaveNet were also suggested for the task~\cite{soni2018time,rethage2018wavenet,fu2019metricgan}.

A different method to improve speech quality is by using speech Bandwidth Extension (BWE) algorithms. In BWE, one is interested in increasing the sampling rate of a given speech utterance. Early attempts were based on Gaussian Mixture Models and Hidden Markov Models~\cite{park2000narrowband,chen2004hmm}, and, more recently, deep learning methods were suggested for the task~\cite{sautter2019artificial,hao2020time}.

Despite the success of the BWE methods, these were mainly applied to lower sampling rates (e.g. 8kHz, 16kHz). Recently, several studies suggested Audio Super Resolution algorithms for upsampling to higher sample rates (e.g. 22.5kHz, 44.1kHz). The authors in~\cite{li2019speech} introduce an end-to-end GAN based system for speech bandwidth extension for use in downstream automatic speech recognition. The authors in~\cite{wang2018speech} suggest to use a WaveNet model to directly output a high sampled speech signal while the authors in~\cite{sheng2019high} suggest using GANs to estimate the mel-spectrogram and then apply a vocoder to generate enhanced waveform.

Unlike previous BWE methods that mainly use generative models for a sequence to sequence mapping, in the waveform or the spectrum domains, our model is conditioned on several high-level features. The proposed model utilizes a speech synthesis model~\cite{gritsenko2020spectral}, an ASR model~\cite{wav2letter}, a Pitch extraction model~\cite{yaapt}, and a loudness feature.

Features extracted from an ASR network were utilized in~\cite{ttsskins} for the task of voice conversion. In addition to differences that arise from the different nature of the task, there are multiple technical differences between the approaches: the generator of~\cite{ttsskins} is autoregressive, the previous method did not include an identity network and modelled new speakers inaccurately, and lastly the loss terms used to optimized the models were different.

\begin{figure*}[h!]
\centering
\includegraphics[scale=0.48,trim={500 82 600 40}]{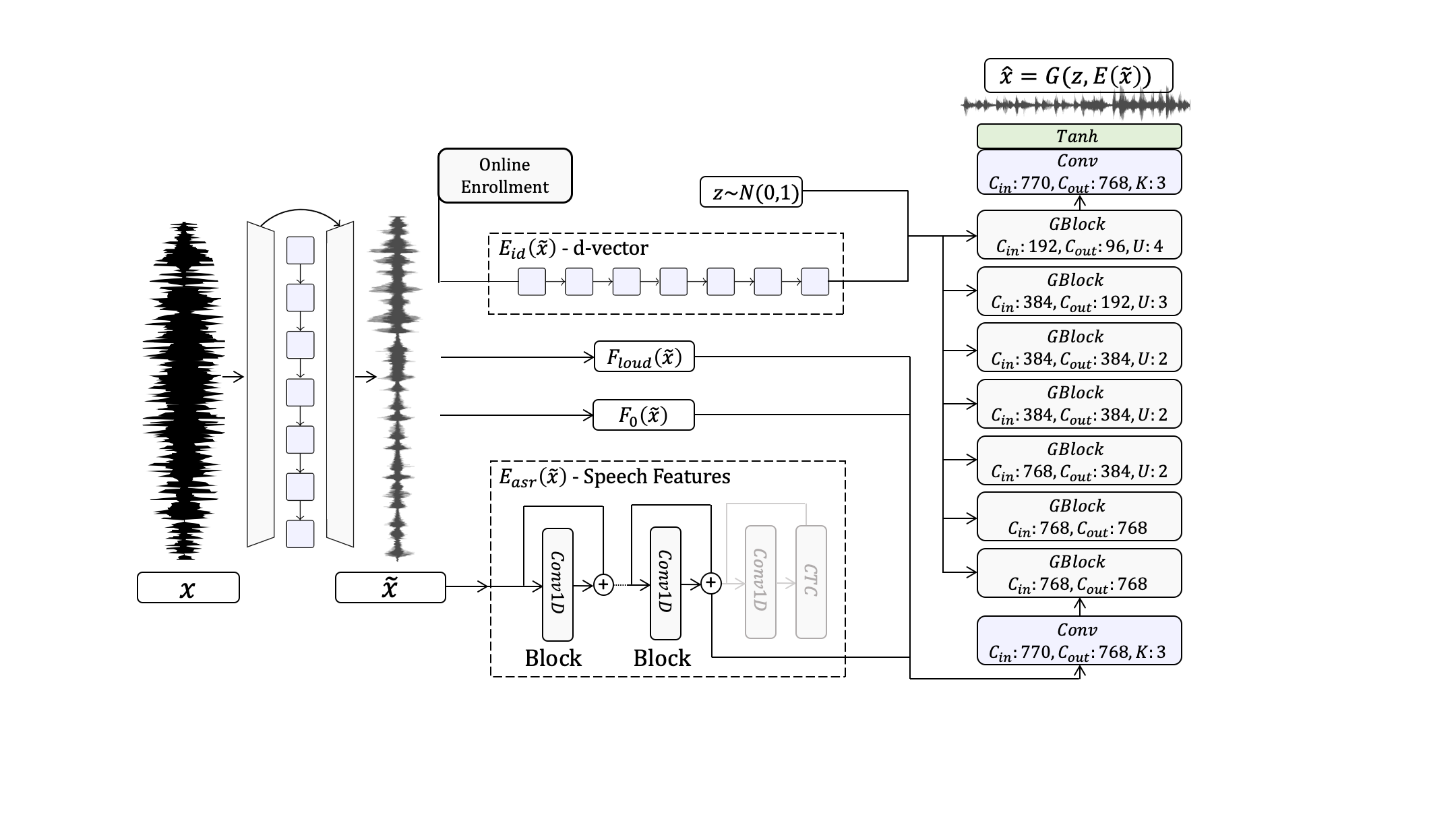}
\caption{\textbf{The proposed speech regeneration architecture.} Noisy speech signal, $\vx$, is initially enhanced using a background removal method. The enhanced signal, $\tilde{\vx}$, is decomposed into four components: (i) Speech features, $E_{asr}(\tilde{\vx})$, using a pretrained ASR encoder, (ii) Fundamental frequency values, $F_0(\tilde{\vx})$ which provide prosody features, (iii) Loudness features $F_{loud}(\tilde{\vx})$, and (iv) d-vector, $E_{id}(\tilde{\vx})$ which embeds the speaker identity. The speech, pitch and loudness features are upsampled across the temporal domain and concatenated. Decoder $G$ receives as input the concatenated signal and conditioned on the concatenation of the identity vector with a random noise vector sampled from a normal distribution. Finally, $G$ outputs the regenerated speech.} 
\label{fig:arch}
\vspace{-0.3cm}
\end{figure*}

\vspace{-0.2cm}
\section{Speech regeneration}
\label{sec:typestyle}
\vspace{-0.1cm}
Our regeneration pipeline is an encoder-decoder network. First, the raw input speech is passed through a background removal method that masks out non-vocal audio. The output is then passed through several subnetworks to generate disentangled speech representations. The output of these subnetworks together with the spectral features condition the speech generative decoder network, which synthesizes the final output. The architecture is depicted in Figure~\ref{fig:arch}.

\subsection{Speech regeneration pipeline} 
Denote the domain of audio samples by $\Xc \subset \R$. The representation for a raw noisy speech signal is therefore a sequence of samples $\vx = (x_1,\ldots, x_T)$, where  $x_t\in\Xc$ for all $1\leq t \leq T$, where $T$ varies for different input sequences. We denote by $\X^*$ the set of all finite-length sequences over $\X$. Consider a single channel recordings with additive noise as follows: $\vx = \vy *\vech + \vn$, where $\vy$ is the clean signal, $\vech$ is the Acoustic Transfer Function, $\vn$ is a non stationary additive noise in an unknown Signal-to-Noise Ratio (SNR), and $*$ is the convolution operator.

Given a training set of $n$ examples, $\Sc = \{(\vx_i, \vy_i)\}_{i=1}^n$, we first remove background noise using a pre-trained state-of-the-art speech enhancement model~\cite{defossez2020real}. Denoting the denoised signal as $\tilde{\vx}$, we define the set of such samples as $\tilde{\Sc} = \{(\tilde{\vx}_i, \vy_i)\}_{i=1}^n$.

Our encoding extracts several different representations in order to capture content, prosody, and identity features separately. Specifically, given an input signal $\tilde{\vx}$, the content representation is extracted using a pre-trained ASR network, $E_{asr}$. In our implementation, we use the public implementation~\cite{jasper} of Wav2Letter~\cite{wav2letter}. The identity representation is obtained in the form of d-vectors~\cite{wan2018generalized}, using an identity encoder $E_{id}$. The d-vector extractor is pre-trained on the VoxCeleb2~\cite{voxceleb2} dataset, achieving a 7.4\% EER for speaker verification on the test split of the VoxCeleb1 dataset. The average activations from the penultimate layer forms the speaker representation. %
Lastly, the prosody representation includes both the fundamental frequency and a loudness feature. The former $F_0(\tilde{\vx})$, is extracted using YAAPT~\cite{yaapt}, which was found to be robust against input distortions. The loudness measurement of the signal, $F_{loud}(\tilde{\vx})$ is extracted using A-weighting of the signal frequencies. The F0 and the Loudness feature are upsampled and concatenated to form the prosody conditioning signal. 

To summarize, the encoding of a denoised signal is given as $E(\tilde{\vx}) = [E_{id}(\tilde{\vx}), E_{asr}(\tilde{\vx}), F_0(\tilde{\vx}), F_{loud}(\tilde{\vx})]$.

\subsection{Learning objective}
The generative decoder network is optimized using the least-squares GAN~\cite{mao2017least} where the decoder, $G$ and discriminator $D$ minimize the following objective,
\begin{equation}
\label{eq:discriminator}
\begin{aligned}
    &L_{adv}(D, G, \tilde{\Sc}) = \sum_{\tilde{\vx} \in \tilde{\Sc}}|| 1 - D(\vxh) ||_2^2 \\
    &L_D(D, G, \tilde{\Sc}) = \sum_{\tilde{\vx} \in \tilde{\Sc}}{[|| 1 - D(\tilde{\vx}) ||_2^2 + || D(\vxh) ||_2^2]}, 
\end{aligned}
\end{equation}
where $\vxh= G(\vz, E(\tilde{\vx}))$, is the synthesized audio sample from a random noise vector sampled from a normal distribution $\vz \sim N(0,1)$. 

The decoder, $G$, is additionally being optimized with a spectral distance loss using various FFT resolutions between the decoder output, $\vxh$, and the target clean signal, $\vy$, as suggested in~\cite{parallelwavegan}. For a single FFT scale, $m$, the loss component is defined as follows:
\begin{equation}
    \label{eq:stft}
    L_{\text{spec}}^{(m)}(\vy, \vxh) = \frac{\| S(\vy) - S(\vxh) \|_F}{\| S(\vy) \|_F} + \frac{\| \log S(\vy) - \log S(\vxh) \|_1}{N}, 
\end{equation}
\noindent where $\| \cdot \|_F$ and $\| \cdot \|_1$ denotes the Forbenius and the $L_1$ norms, $S$ is the Short-time Fourier transform (STFT), and $N$ the number of elements. 

The multi-scale spectral loss is achieved by summing Equation (\ref{eq:stft}) over the following FFT resolutions $M=[2048,1024,512,256,128,64]$ as follows,
\begin{equation}
    L_{spec}(\vy, \vxh) =  \frac{1}{|M|}\sum_{m \in M} L_{spec}^{(m)}(\vy, \vxh).
\end{equation}

Inspired by the recently suggested spectral energy distance formulation of the spectral loss~\cite{gritsenko2020spectral}, the spectral loss is applied as part of a compound loss term:
\begin{equation}
\begin{aligned}
    L_{sed}(G, \tilde{\Sc}) = \sum_{(\tilde{\vx}, \vy) \in \tilde{\Sc}}  L_{spec}&(\vy, G(\vz_1, E(\tilde{\vx})))\\
    + & L_{spec}(\vy, G(\vz_2, E(\tilde{\vx}))) \\ 
    - & L_{spec}(G(\vz_1, E(\tilde{\vx})), G(\vz_2, E(\tilde{\vx}))),
\end{aligned}
\end{equation}
where $\vz_1, \vz_2$ are two different normally distributed random noise vectors. Intuitively, the energy loss maximizes the discrepancy between different values of $\vz$.

Replacing $ L_{spec}$ with $ L_{sed}$ has improved the quality of the generated audio. Specifically, in our experiments, we noticed the removal of metallic effects from the generated audio.

Overall, the objective function of the decoder generator, $G$, is defined as: 
\begin{equation}
\label{eq:generator}
\begin{aligned}
L_g(G, D, \tilde{\Sc}) = L_{sed}(G, \tilde{\Sc}) + \lambda \cdot L_{adv}(D, G, \tilde{\Sc}),
\end{aligned}
\end{equation}
where $\lambda$ is a tradeoff parameter set to 4 in our experiments.

The preliminary enhancement of the input signal, $\vx$, produces the denoised signal, $\tilde{\vx}$, sampled at 16kHz. The decoder then receives as input the concatenated and upsampled conditioning signal, $E(\tilde{\vx})$, sampled at 250Hz. $G$ is conditioned on the concatenation of the noise vector, $\vz$, and the speaker identity, $E_{id}(\tilde{\vx})$, while the input to the model is $E_{asr}(\tilde{\vx})$ concatenated with $F_0(\tilde{\vx})$ and $F_{loud}(\tilde{\vx})$. Finally, the proposed model outputs a raw audio signal sampled at 24kHz. %

The architecture of the decoder $G$ is based on the GAN-TTS~\cite{gantts} architecture consisting seven GBlocks. A GBlock contains a sequence of two residual blocks, each with two convolutional layers. The convolutional layers employ a kernel size of 3 and dilation factors of $1, 2, 4,  8$ to increase the network receptive field. Before each convolutional layer, the input is passed through a Conditional Batch Normalization~\cite{dumoulin2016learned}, conditioned on a linear projection of the noise vector and speaker identity, and a ReLU activation. The final five GBlocks upsample the input signal by factors of $2, 2, 2, 3, 4$ accordingly, to reach the target sample rate. Figure~\ref{fig:arch} includes the hyperparameters of each GBlock.

While GAN-TTS~\cite{gantts} was originally trained with both conditional and unconditional discriminators operating in multiple scales, we found in preliminary experiments that the proposed method can generate high-quality audio using a single unconditional discriminator, $D$. Moreover, our architecture of $D$ is much simpler than the one proposed in previous work. It involves a sequence of seven convolutional layers followed by a leaky ReLU activation with a leakiness factor of 0.2, except the final layer. The number of filters in each layer is $16, 64, 256, 1024, 1024, 1024, 1$ with kernel sizes of $15, 41, 41, 41, 41, 5, 3$. Finally, to stabilize the adversarial training, both the discriminator and the decoder employ spectral normalization~\cite{miyato2018spectral}.

\vspace{-0.2cm}
\section{Evaluation}
\vspace{-0.1cm}
\label{sec:Experiments}
We present a series of experiments evaluating the proposed method using both objective and subjective metrics. We start by presenting the datasets used for evaluation. Next, a comparison to several competitive baselines is held, and we conclude with a discussion on the computational efficiency of the proposed method. 

\vspace{-0.2cm}
\subsection{Datasets}
\label{ssec:Datasets}
\vspace{-0.1cm}

We evaluated our method for speech regeneration on two different datasets. The first one is the Device and Produced Speech (DAPS)~\cite{daps} dataset. DAPS is comprised of twelve different recordings for each sample; two devices and seven acoustic environments. For a fair comparison, we used the same test partition employed by HifiGAN~\cite{su2020hifi}. For this benchmark, we optimized the generator on the public VCTK~\cite{vctk} dataset, which consists of 44 hours of clear professionally recorded voices of 109 speakers.

The second dataset is a standard benchmark in the speech enhancement literature~\cite{valentini}. The dataset contains clean and artificially added noisy samples. The clean samples are based on the VCTK dataset~\cite{vctk}, where the noises are sampled from~\cite{thiemann2013demand}. The dataset is split into a predefined train, validation, and test sets. In both settings, we resample the audio samples to 24kHz.

\begin{table}[t!]
\centering
\caption{MUSHRA and perceptual metrics on the DAPS dataset~\cite{daps}}
\label{tab:daps}
\begin{tabular}{lccccccc}
\toprule

Method & MUSHRA~$\uparrow$ & FDSD~$\downarrow$ & cFDSD~$\downarrow$ \\
\midrule
Clean           &  4.88$\pm$0.36 &  --   & -- &\\
\midrule
Noisy           &  2.62$\pm$0.96 &  11.31   & 7.79 &\\
WaveUNet~\cite{stoller2018wave}         &  2.09$\pm$1.14 &  13.09   & 10.16 \\
MetricGAN~\cite{fu2019metricgan}        &  2.92$\pm$0.92 &  8.23    & 4.90 \\
HiFiGAN~\cite{su2020hifi}               &  3.55$\pm$1.07 &  7.50    & 3.53 \\
Demucs~\cite{defossez2020real}          &  3.13$\pm$1.13 &  10.36   & 3.72 \\
\midrule
\bf Ours   &  \bf 3.76$\pm$1.02 &  \bf 6.61    & \bf 3.16 \\

\bottomrule
\end{tabular}
\vspace{-.3cm}
\end{table}

\vspace{-0.2cm}
\subsection{Experimental results}
\vspace{-0.1cm}
\label{ssec:Evaluation}

\noindent{\bf Evaluation Metrics.\quad}
We evaluated our method using both objective and subjective metrics. For subjective metric, we used MUSHRA~\cite{mushra} -- We ask human raters to compare samples created from the same test signal. The clean sample is presented to the user before the processed files and is labeled with a 5.0 score. 

{For objective metrics we used two distances proposed in~\cite{gantts}: (i) Fr\'echet Deep Speech Distance (FDSD) - a distance measure calculated between the activations of two randomly sampled sets of output signals to the target sample using DeepSpeech2~\cite{deepspeech2} ASR model; (ii) Conditional Fr\'echet Deep Speech Distance (cFDSD) - similar to FDSD, but computed between the generated output to its matched clean target. Unlike cFDSD, where the generated output should match the target signal, in FDSD the random sets do not have to match to the target utterance.  Note that the Fr\'echet distances are computed using a different ASR network than the one used for conditioning the proposed model.} 

{We do not employ the PESQ metric~\cite{pesq}, which was designed to quantify degradation due to codecs and transmission channel errors. PESQ validity for other tasks is questionable, e.g., it shows a low correlation with MOS~\cite{reddy2019scalable}. Moreover, it is defined for narrowband and wideband only. %
}

\smallskip
\noindent{\bf Results.\quad}
Table~\ref{tab:daps} shows the performance of our method on the DAPS dataset. We compared our regeneration method to several competitive baselines in the domain of speech enhancement. As can be seen, our method outperforms the baselines in all metrics. The regenerated speech has substantially lower perceptual distances, and scored convincingly higher on the MUSHRA test compared to the baselines.

Results for the noisy VCTK dataset are presented in Table~\ref{tab:vctk}. Our method is superior to the baselines models on the noisy VCTK as well, however, improvements are modest compared to DAPS. This is due to VCTK being a less challenging dataset for enhancement tasks. Unlike DAPS, which its test split was recorded in real noisy-reverberant environments, in VCTK noise files were added to the clean samples to artificially generate noisy samples. %

\begin{table}[t!]
\centering
\caption{MUSHRA and perceptual metrics on the noisy VCTK dataset~\cite{valentini}.}
\label{tab:vctk}
\begin{tabular}{lccccccc}
\toprule

Method & MUSHRA~$\uparrow$ & FDSD~$\downarrow$ & cFDSD~$\downarrow$ \\%fix this line
\midrule
Clean                               &  4.74$\pm$0.47 & -- & -- \\
\midrule
Noisy                               &  2.74$\pm$0.85 &  4.72 & 1.79 \\
HifiGAN~\cite{su2020hifi}           &  3.92$\pm$0.88 &  4.34 & 0.91 \\
Demucs~\cite{defossez2020real}      &  3.16$\pm$0.80 &  4.31 & 0.77 \\
\midrule
\bf{Ours}                           &  \bf 4.00$\pm$0.89 & \bf 4.22 & \bf 0.73 \\
\bottomrule
\end{tabular}
\vspace{-0.3cm}    
\end{table}

\vspace{-0.2cm}
\subsection{Computational efficiency}
\vspace{-0.1cm}
\label{sec:efficiency}

{The accessibility to speech enhancement greatly relies on its efficiency and ability to be applied while streaming. We have efficiently implemented a server-based module using PyTorch JIT that is able to fetch speech audio and regenerate it with a \emph{Real-Time Factor} of 0.94. All modules required to compute the conditioning vector run in parallel either on the V100 NVIDIA GPU or Intel Xeon CPU E5.  The final pipeline currently has a latency of about 100 milliseconds (ms), which takes into account the receptive field, future context and computation time of every module. All modules operate at a receptive field of up to 40ms. The ASR network and the denoiser use a future context of 20ms and 32ms respectively. The rest of the modules require no future context and are trained to be fully causal. Such latency can fit Voice-Over-IP applications.
}

\vspace{-0.4cm}
\section{Conclusions}
\vspace{-0.2cm}

\label{sec:Discussion}
We present an enhancement method that goes beyond the limitations of a given speech signal, by extracting the components that are essential to the communication and recreating audio signals. The method is taking advantage of recent advances in ASR technology, as well as in pitch detection and identity-mimicking TTS and voice conversion technologies.

The recreation approach can also be applied when one of the components is manipulated. For example, when editing the content, modifying the pitch in post-production, or replacing the identity. An interesting application is to create super-intelligible speech, which would enhance the audience's perception and could be used, e.g., to improve educational presentations.

\bibliographystyle{IEEEbib}
\bibliography{regen}

\end{document}